# New HI views of the Galaxy and the Magellanic Clouds

**Snežana Stanimirović**

University of WisconsinMadison, Department of Astronomy, 475 N Charter St, Madison, WI 53703, USA

**Abstract.** Atomic hydrogen (HI) is a vital player in the star-formation process in galaxies. It is the raw fuel for making molecules, an important shielding agent against interstellar radiation, and a buffer that soaks up the energy and momentum of stellar feedback. While for many years detailed studies of the HI thermal structure have been possible only in the Milky Way, the SKA pathfinders are expanding our view beyond the Solar neighborhood allowing for crucial tests of the HI heating and cooling processes under a wide range of physical conditions. This overview article highlights a few recent results and emphasizes areas where future observations can make large contributions. We show that the cold HI disk of the Milky Way is extended and flared, yet appears spatially coupled to the molecular gas. The cold neutral medium (CNM) in the Milky Way is colder and more abundant at higher optical extinctions due to more intense cooling and shielding. A comparison between the Milky Way, the Small Magellanic Cloud, the Large Magellanic Cloud and NGC 6822 shows a good agreement with predictions from recent numerical simulations on how the CNM fraction depends on metalliciity. The fraction and spatial distribution of the thermally unstable HI remain as open questions and observationally have not been studied beyond the Milky Way. The excitation temperature of the warm neutral medium (WNM) is not well understood. Observations suggest higher WNM temperatures than what is seen in numerical simulations. The SKA, once operational, will take the HI studies into a new era of dense HI absorption grids for the Milky Way and nearby galaxies, and the ability to study the HI thermal structure beyond the Local Group.

---

## 1. Introduction

The diffuse interstellar medium (ISM), traced by neutral hydrogen (HI), is a vital player in the star formation process and represents a major contribution to the overall ISM's mass. The HI is the raw fuel for making molecules, and provides shielding against interstellar radiation. The HI exists in galaxies over a wide range of temperatures, $\sim 10$ to $\sim 10^4$ K. However, only densest and coldest HI – the cold neutral medium (CNM) – is expected to go into making molecular hydrogen ($H_2$) on the surface of dust grains (Spitzer and Jenkins 1975; Elmegreen 1993; Krumholz et al. 2009). This theoretical expectation has been confirmed observationally (Rybarczyk et al. 2022). As a consequence, recent numerical simulations show a strong correlation between spatial distributions of the CNM and the star formation rate in simulated galaxies, calling this the "CNM Kennicutt-Schmidt" relation (Smith et al. 2023). The HI in galaxies is subjected to strong stellar feedback and carries imprints of stellar activity in the form of numerous discrete structures, e.g. shells, chimneys, and the overall level of interstellar turbulence.

Theoretical models of steady-state ISM heating and cooling predict the thermal structure of the HI gas. In the Solar neighborhood, two thermally stable phases — the cold neutral medium (CNM) and warm neutral medium (WNM) — are expected, with the kinetic temperature ($T_k$) being mainly in the range $\sim 50$ K to 250 K and 5000 K to 8000 K, respectively (e.g. Wolfire et al. 2003). These steady-state models demonstrate that HI in the range of $\sim 250\ \mathrm{K} \leq T_k \leq$ 5000 K is thermally unstable and short-lived, although this depends on the details of the local heating and cooling processes (Field et al. 1969; Wolfire et al. 1995; Wolfire et al. 2003; Bialy and Sternberg 2019). The local conditions of the ISM influence the properties of distinct HI phases. For example, in metal-poor gas with low dust content, the photoelectric heating from the surface of dust grains and polycyclic aromatic hydrocarbons (PAHs) is less effective (Bialy and Sternberg 2019). As a result, at low metallicity theoretical studies expect that the CNM should be colder and less abundant. Additionally, when the UV radiation field increases by a factor of 10, the thermal equilibrium curve, which marks where thermally-stable CNM and WNM can co-exist, shifts to higher pressures resulting in the CNM properties resembling those of molecular clouds in the Milky Way (Dickey et al. 2000; McClure-Griffiths et al. 2023). However, observational constrains on the HI thermal structure currently come exclusively from the Solar neighborhood.

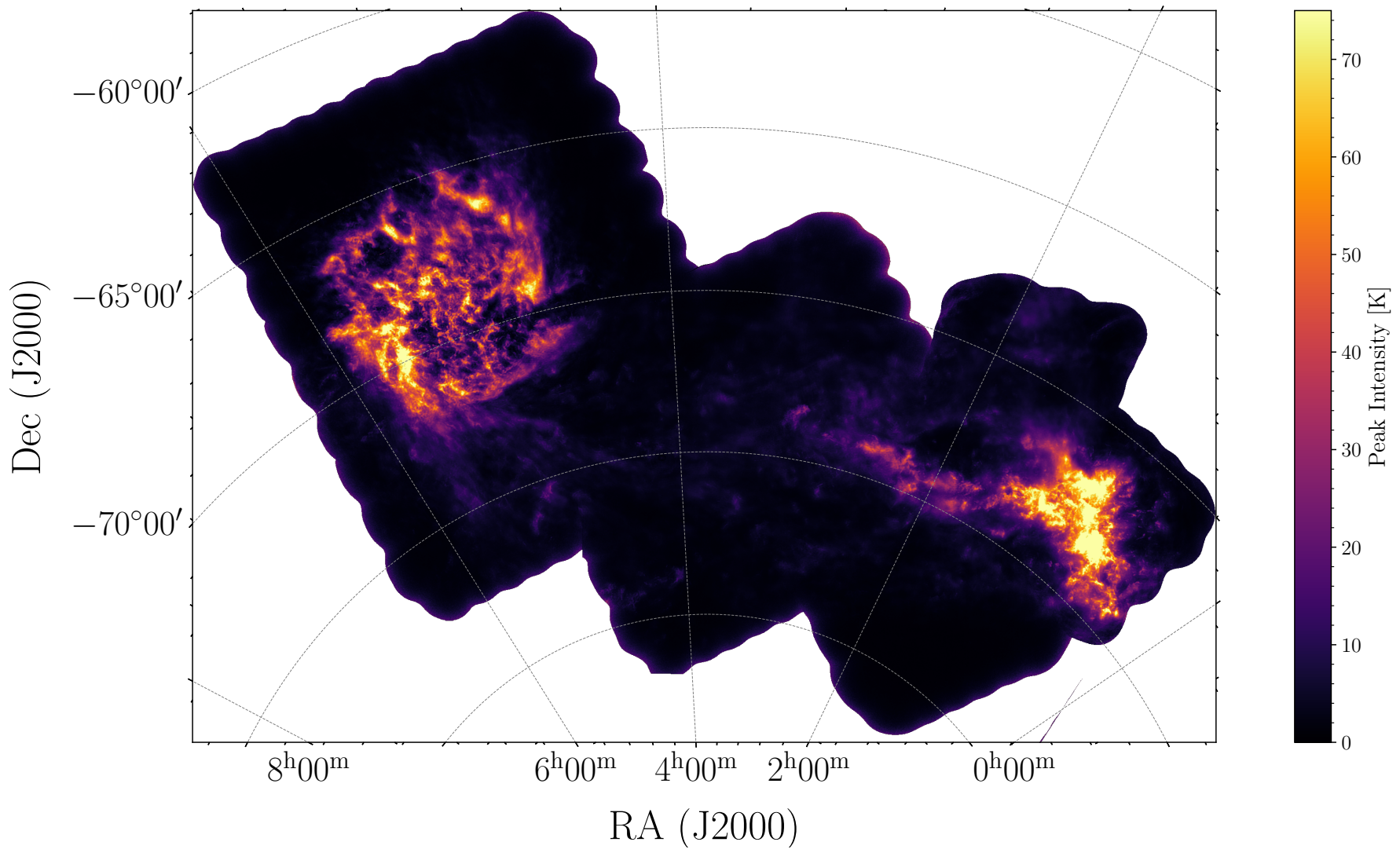

**Figure 1.** The peak brightness temperature image of the SMC, LMC and the Bridge obtained by the GASKAP-HI collaboration. Courtesy Jackie Ma and Nickolas Pingel.

The 21-cm line of atomic hydrogen has been the key tracer of the neutral ISM in the Milky Way, and galaxies in general, since its discovery by Ewen and Purcell (1951). However, for many years detailed studies of the HI structure and thermodynamics have been exclusively possible only in the Milky Way where high resolution and sensitivity observations can be obtained, with the Small Magellanic Cloud (SMC) and the Large Magellanic Cloud (LMC) providing only a glimpse of how HI properties depend on interstellar environment. With their metallicity of 1/5 for the SMC (Dufour 1984) and 1/2 Solar for the LMC (de Grijs et al. 2014), the Magellanic Clouds have served as key windows into physical processes under conditions similar to those of galaxies in the early Universe. The wealth of recent radio telescopes and observations brings HI studies to a pivotal moment where, for the first time, the line between ISM studies in the Milky Way and those of nearby galaxies is starting to blur. The detailed ISM studies that until recently were possible only in the Milky Way and occasionally in the Magellanic Clouds, can now be routinely done in nearby galaxies with interstellar conditions very different from what we find in the Solar neighborhood. This allows us to greatly expand the parameter space and test theoretical and numerical models over a much more diverse range of interstellar conditions.

There have been several articles summarizing what we know about the HI properties in the Milky Way that readers should consult for comprehensive reviews, e.g. Kulkarni and Heiles (1987); Burton (1988); Dickey and Lockman (1990); Kalberla and Kerp (2009), and more recently McClure-Griffiths et al. (2023). In this overview article, we primarily highlight a few recent results and emphasize areas where future observations can make large contributions. Specifically, in Section 1 we provide a short summary of the most commonly used HI emission and absorption surveys over the last 10-15 years. We then start with a discussion of the large-scale properties of the Milky Way's warm and cold disks in Section 3. Most of the article is highlighting what we know about how HI properties respond to the underlying physical conditions, Section 4. Section 5 focuses on the HI mass fraction in thermally stable and unstable phases and the spatial distribution of the CNM. We summarize key ideas in Section 6.

## 2. HI surveys

### 2.1. *HI emission surveys*

Many HI emission surveys of the Milky Way over the last 20-30 years have shown complex HI structure over a wide range of spatial scales. While a complete survey summary can be found in McClure-Griffiths et al. (2023), in Table 1 we summarize few recent surveys that are still commonly used to emphasize the range of angular scales typically probed and survey complementarity.

In the case of the Magellanic Clouds, large mosaics with the Australia Telescope Compact Array (ATCA) from the late 1990s, combined with single-dish data from the Parkes radio telescope, served as key HI resources for over 25 years. Only very recently these data sets are being superseded with high resolution observations from GASKAP. While more sensitive, and even higher resolution observations will come in the near future from GASKAP, the new data sets are already enabling a wide array of science applications. Figure 1 shows an HI peak brightness temperature image of the SMC, LMC and the Bridge in between at 30″ resolution. The emission surveys show the spatial distribution and kinematics of the HI gas, provide a lower limit on the HI mass (under the assumption of low optical depth), and the upper limit on the kinetic temperature based on the emission line widths.

**Table 1.** A comparison of common HI emission surveys.

| Survey | Telescope | Area | Resolution | Velocity resolution | Sensitivity |
|---|---|---|---|---|---|
| **Milky Way** | | | | | |
| GALFA-HI[1] | Arecibo | 32% | 4.0′ | $0.2\ \mathrm{km\,s^{-1}}$ | 0.3 K |
| HI4PI[2] | Effelsberg + Parkes | 100% | 16′ | $1.5\ \mathrm{km\,s^{-1}}$ | 0.05 K |
| THOR[3] | VLA | $<1\%$ | 18″ | $1.5\ \mathrm{km\,s^{-1}}$ | 16 K |
| GASKAP[4] | ASKAP | ongoing | 30″ | $0.5\ \mathrm{km\,s^{-1}}$ | $\sim 1.5$ K |
| **Magellanic Clouds** | | | | | |
| SMC[5] | ATCA + Parkes | SMC | 98″ | $1.65\ \mathrm{km\,s^{-1}}$ | 1.3 K |
| LMC[6] | ATCA + Parkes | LMC | 1′ | $1.65\ \mathrm{km\,s^{-1}}$ | 2.5 K |
| SMC[7] | ASKAP + Parkes | SMC | 30″ | $0.98\ \mathrm{km\,s^{-1}}$ | 1.1 K |
| LMC[8] | ASKAP + Parkes | LMC | 30″ | $0.98\ \mathrm{km\,s^{-1}}$ | $\sim 1$ K |

*References*: (1) Peek et al. (2011), (2) HI4PI Collaboration et al. (2016), (3) Beuther et al. (2016), (4) Dickey et al. (2013); Pingel et al. (2022), (5) Stanimirovic et al. (1999), (6) Kim et al. (2003), (7) Pingel et al. (2022) , (8) GASKAP collaboration, in preparation

### 2.2. *HI absorption surveys*

The most direct way of studying the HI thermal structure, and detecting the cold HI, is through the 21-cm absorption against strong radio continuum sources (Murray et al. 2018; Jameson et al. 2019; Dempsey et al. 2022). Such measurements enable constraints of the excitation temperature (the same as the kinetic temperature at high densities) and optical depth when the associated HI emission spectrum is also considered. The last two decades have seen an explosion of HI absorption surveys of the Milky Way (Stanimirović et al. 2014; Murray et al. 2021; Roy et al. 2013; Wang et al. 2020; Basu et al. 2022), with some of the key studies being the Millennium survey (Heiles and Troland 2003), the Giant Metrewave Radio Telescope survey (Mohan et al. 2004a,b), 21-SPONGE Murray et al. (2015, 2018), and the HI absorption in the direction of Taurus, California, Rosette, Mon OB1, NGC 2264 (Nguyen et al. 2019). Key data products from many of these surveys have been compiled into a database BIGHICAT by McClure-Griffiths et al. (2023). This database contains over 1000 unique Gaussian components and is enabling exploration of a wide range of science questions. As the power of large data sets is clearly evident (see other sections in this paper), we encourage the community to continue to contribute to the BIGHICAT data set. While largely focusing on the Magellanic System so far, the GASKAP-HI survey has provided dense grids of HI absorption spectra for the foreground regions. These are among the densest samples of HI absorption ever detected, with a source density of 11 deg$^{-2}$ (Nguyen et al. 2024), and enabling new science (see Nguyen's and Lynn's contributions in this volume).

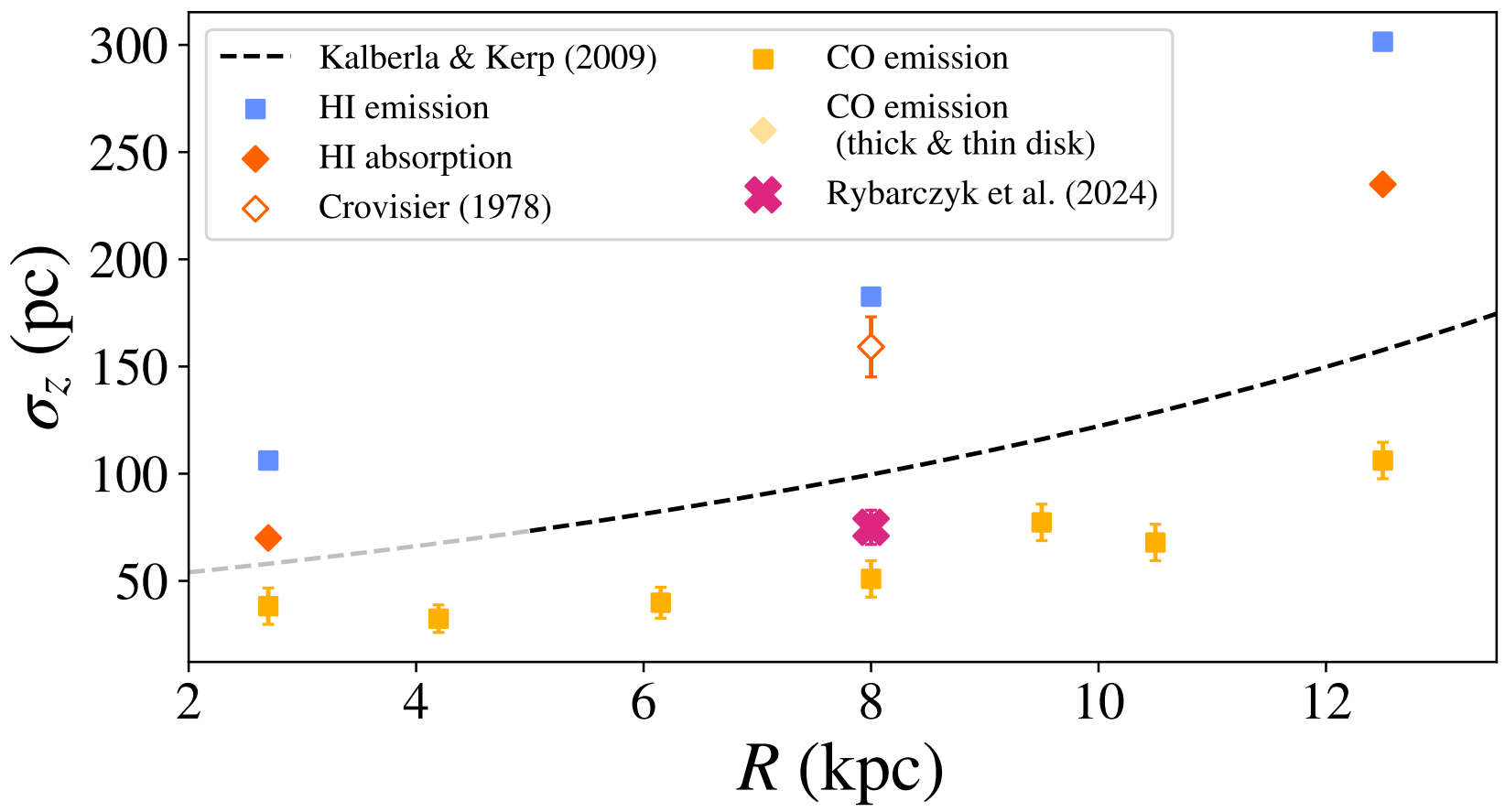

**Figure 2.** The scale height of gas as a function of Galactocentric radius from Rybarczyk et al. (2024). Yellow squares indicate the thickness of the molecular gas disk derived from observations of CO in emission based on data and references in Heyer and Dame (2015). Orange diamonds indicate the thickness of the cold HI disk derived from observations of HI in absorption from Dickey et al. (2009) and Dickey et al. (2022). The Solar neighborhood estimate for the cold HI disk from Crovisier (1978) is shown as an unfilled orange diamond. Blue squares indicate the thickness of the total HI disk derived from observations of HI in emission (HI4PI Collaboration et al. 2016). The new estimate from Rybarczyk et al. (2024) is shown as a red X. A black dashed line shows the exponential fit to the HI disk thickness from Kalberla and Kerp (2009), valid for Galactocentric radii 5 - 35 kpc.

In the case of the Magellanic Clouds, there have been several recent targeted studies of HI absorption (Jameson et al. 2019). However, the GASKAP-HI survey of the Magellanic Clouds has resulted in the largest ever samples of HI absorption spectra. For example, in Dempsey et al. (2022) HI absorption was detected in the direction of 65 of 229 background radio sources. For comparison, the Dickey et al. (2000) study using the ATCA had a sample of 13 detected absorption lines only. In the case of the LMC over 222 (out of 1567 sources) HI absorption detections have been made (Chen, Steffes et al, in preparation). A lot more will come in the near future with the upcoming deep surveys of the SMC and LMC by ASKAP. The wide-field surveys are clearly revolutionizing HI absorption studies.

## 3. Milky Way's warm and cold HI disks

The HI emission surveys of the Milky Way have traced a neutral disk that extends all the way to $\sim 60$ kpc from the Galactic center (Kalberla and Kerp 2009; McClure-Griffiths et al. 2023).

The HI surface density within the disk is roughly constant up to $\sim 10$ kpc, with an exponential decline beyond this radius. Beyond about 30 kpc, it is expected that the HI disk has low density and is likely clumpy (Kalberla and Dedes 2008). Since early maps of the Milky Way HI, it has been known that the HI disk seen in emission is warped and flared, e.g. Dickey and Lockman (1990). The large-scale reconstructions of the outer HI disk, by inverting HI position-position-velocity data cubes into a 3D spatial structure have suggested interesting disk features, with prominent scalloping in the extreme outer parts (Levine et al. 2006; Koo et al. 2017; Mertsch and Phan 2023). While these methods depend on a detailed implementation of the HI rotation curve and radial motions, they show promise for mapping detailed disk structure in the near future.

The structure and extent of the cold HI disk have been less understood. The pioneering studies of Strasser et al. (2007) and Dickey et al. (2009) showed the presence of HI absorption all the way to 18-25 kpc from the Galactic center and that the cold HI is mostly found along spiral arms. Recent observations with GASKAP by Dickey et al. (2022) enabled stacking of HI absorption spectra based on the terminal velocity – longitude relationship from McClure-Griffiths and Dickey (2016). The stacked spectra showed the presence of HI absorption at 20-40 kpc from the Galactic center. The width of the stacked absorption spectra significantly changes from 8 to 40 kpc, suggesting that the cold HI disk flares, just like the warm HI disk, at large Galactocentric radii.

We can now compare the thickness of the cold HI disk with what is known about HI emission and the cold molecular (CO) disk of the Milky Way, see Figure 2. This figure was recreated from Figure 11 of McClure-Griffiths et al. (2023), and shows the thickness of the disk as a function of Galactocentric radius measured in HI emission (blue squares), HI absorption (orange diamonds), and CO emission (yellow squares). Orange diamonds indicate the thickness of the cold HI disk derived from observations of HI in absorption (Dickey et al. 2009; Dickey et al. 2022). Blue squares indicate the thickness of the total HI disk derived from observations of HI in emission (HI4PI Collaboration et al. 2016). Until recently, the only constraint on the cold disk thickness in the Solar neighburhood was from Crovisier (1978), shown as an open diamond, and the general understanding has been that the cold HI disk scale height increases from $\sim 50$ pc at $\sim 3$ kpc, to $\sim 120$ pc at 8 kpc , and then all the way to $\sim 300$ pc at $\sim 12$ kpc.

However, a recent study by Wenger et al. (2024) showed that the method used by Crovisier (1978) over-estimated the disk scale height due to improper handling of non-uniform sampling of data. Further, Rybarczyk et al. (2024) showed that at $R = 8$ kpc, the cold HI scale height is only $\sim 75$ pc, in essence staying relatively flat across the inner Galaxy. This challenges the conventional picture that the vertical thickness gradually rises with Galactocentric radius, e.g. McClure-Griffiths et al. (2023). Furthermore, Rybarczyk et al. (2024) showed that the optically thicker HI sample with $\tau > 0.1$ has a thickness close to that measured for the molecular gas disk, while the optically thinner sample with $\tau < 0.1$ is about halfway between the molecular gas disk and the disk measured in HI emission. These results suggest that cold HI is spatially coupled to the molecular gas.

## 4. HI as a function of local conditions

As discussed in the Introduction, theoretical considerations of heating and cooling processes predict that HI properties (e.g. temperature, cold gas fraction) should reflect changes in local environmental parameters (e.g. metallicity, radiation field, density). By contrasting HI measurements between the Milky Way, the SMC and the LMC, and other nearby galaxies, we have the opportunity to test theoretical predictions and in turn examine different heating and cooling processes.

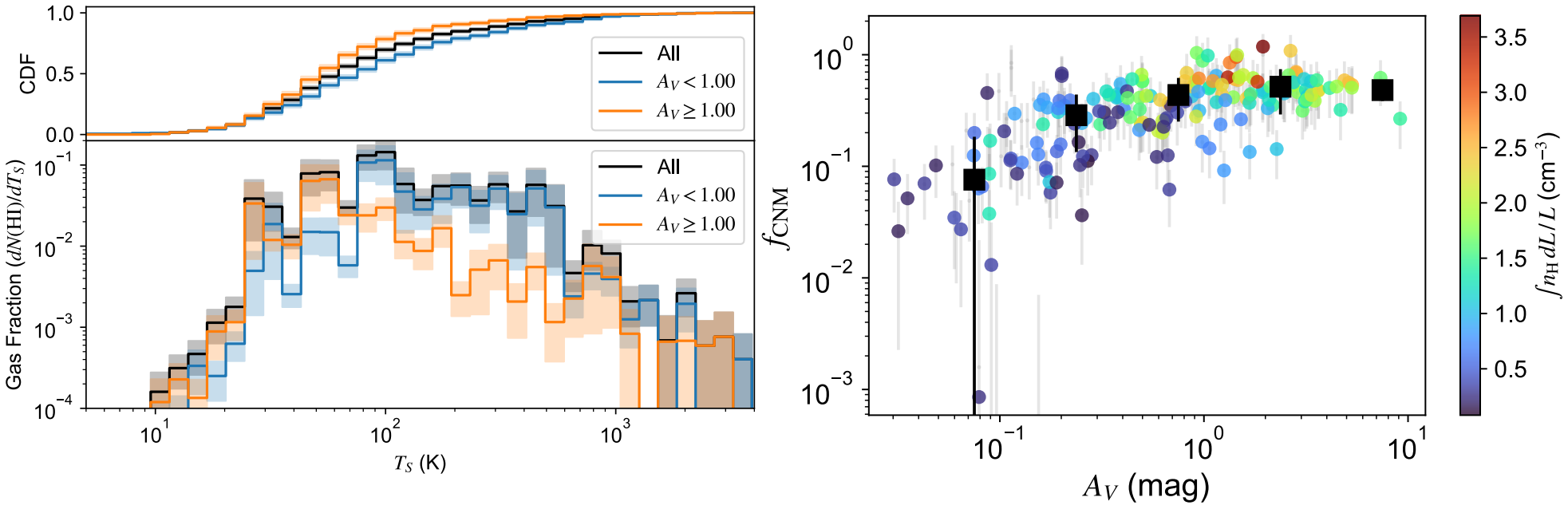

**Figure 3.** (Left) Fraction of the total HI column density detected as a function of spin temperature from the BIGHICAT absorption compilation. The overall fraction (black) is divided basis on the visual extinction, $A_V$. For high extinction environments ($A_V > 1.0$) the temperature range observed in the CNM is narrow and peaked at $T_s \approx 50$ K, whereas the lower extinction environments show a broader range of CNM temperatures peaked closer to $T_s \approx 100$ K. (Right) The CNM fraction from the BIGHICAT dataset. The HI absorption data come from a very sharp sampling, while $A_V$ data are from Planck with angular resolution of $\sim 5'$. The high-$A_V$ limit in this figure is not well defined as HI spectra can be saturated and complex for interpretation. Both figures are from McClure-Griffiths et al. (2023).

### 4.1. *The CNM temperature and fraction as a function of extinction*

While there have been sporadic claims of directions in the Milky Way with a low excitation temperature, e.g. Knapp and Verschuur (1972); Heiles and Troland (2003), systematic investigations of how $T_s$ depends on local conditions have been rare. For example, Strasser et al. (2007) used data from the International Galactic Plane Survey and suggested that $T_s$ remains constant across the Milky Way disk by finding $T_s = 48 \pm 10$ K for the inner Galaxy vs. $T_s = 38 \pm 10$ K for the outer Galaxy.

A large sample of $T_s$ measurements provided by BIGHICAT showed conclusively, for the first time, that $T_s$ depends on local conditions (McClure-Griffiths et al. 2023). As shown in Figure 3 (left), the HI column density (or mass) distribution peaks at different values of $T_s$ depending on optical extinction: the $A_V < 1$ sample peaks at $T_s = 80 - 100$ K, while the $A_V > 1$ sample peaks at $T_s \sim 50$ K. This shows that regions with higher extinction have more intense cooling and shielding, and therefore a lower excitation temperature.

Similarly, the BIGHICAT data were used to show that the cold HI fraction, Figure 3 (right), defined as the ratio of the CNM column density over the total HI column density, $f_{\mathrm{CNM}}$, increases with $A_V$ from $\sim 10\%$ to $\sim 80\%$. The data points in this figure are color-coded based on the mean volume density along the line of sight obtained from the 3D dust data (Lallement et al. 2019) and show that points with higher $A_V$ and higher $f_{\mathrm{CNM}}$, also have high mean densities. The conclusion is that higher local density implies more cooling and shielding, and therefore more CNM (McClure-Griffiths et al. 2023).

### 4.2. *The CNM fraction as a function of Galactocentric radius*

We next consider how the CNM fraction changes across the Milky Way disk. Studies of HII regions have shown that the Milky Way disk has a strong metallicity gradient, e.g. Wenger et al.

(2019). At a Galactocentric radius of 35 kpc, the metallicity drops by a factor of $\sim 10$, reaching values similar to what is found in the SMC. Similarly, at such large Galactocentric radii the interstellar radiation field is expected to be a factor of $\sim 100$ lower than in the Solar neighborhood. For such conditions, theory predicts that the CNM fraction should start to decrease, e.g. Wolfire et al. (2003).

Surprisingly, observations have suggested a relatively constant CNM fraction as a function of Galactocentric radius so far. The optical-depth-weighted mean spin temperature, $\langle T_s \rangle$ , defined as:

$$\langle T_s \rangle = \frac{\int T_B(v) dv}{\int 1 - e^{-\tau(v)} dv} = \frac{T_s}{f_{\rm CNM}} \tag{1}$$

is commonly used as an indicator of the CNM fraction, where $T_s$ is an average cloud temperature. To calculate $\langle T_s \rangle$ we need HI emission ($T_B(v)$) and optical depth ($\tau(v)$) spectra, without separation of individual velocity components. This quantity is therefore frequently used for complex spectra, e.g. within the disk of the Milky Way.

Dickey et al. (2009) compiled HI absorption-emission measurements in the Galactic plane from the three Galactic Plane Surveys (CGPS, SGPS and VGPS) to measure $\langle T_s \rangle$ as a function of Galactocentric radius, and showed that $\langle T_s \rangle \approx 300$ K from $R = 8$ to $R = 25$ kpc. Using GASKAP observations, Dickey et al. (2022) showed that $\langle T_s \rangle$ remains constant all the way to $R \sim 40$ kpc. Based on Equation (1), these results imply a CNM fraction of 15-20% at $R \sim 40$ kpc if a constant temperature of $T_s \sim 50$ K is assumed. While $\langle T_s \rangle$ is the quantity easily obtained from observations, to relate it to $f_{\rm CNM}$ we require the knowledge of an average cloud temperature. As shown in McClure-Griffiths et al. (2023), if the typical cold cloud temperature decreases with Galactic radius, as expected based on theoretical models from Wolfire et al. (2003) with $T_s = 85$ K at $R = 8.5$ kpc to $T_s = 44.1$ K at $R = 18$ kpc, then $f_{\rm CNM}$ also decreases from $\sim 0.3$ near the Solar circle to less than 0.05 at $R \sim 25$ kpc. Essentially, as the heavy element abundance decreases with $R$, the cooling rate also decreases. At the same time the radiation field and the dust abundance, which are key for photoelectric heating, also both decrease with $R$. If the heating rate decreases faster than the cooling rate, the optical-depth-weighted mean spin temperature $\langle T_s \rangle$ will be declining with $R$.

Clearly, observational constraints on how the typical cold cloud temperature varies with Galactocentric radius deserve future studies as this is the key ingredient in explaining $\langle T_s \rangle$ distribution. As we showed in Section 4.1, it is already known that $T_s$ is not constant throughout the Milky Way disk but depends on local conditions. To add more to the ongoing puzzle, recent simulations by Smith et al. (2023) investigated the CNM distribution of an isolated spiral galaxy and found that the CNM fraction was roughly constant across most of the disk and sharply decreased in the outer disk regions. Their explanation is that the falling interstellar radiation field compensates for the decreasing column density at large radii, leaving $\langle T_s \rangle$ relatively constant. Future observational constraints will be crucial to test this prediction.

### 4.3. *The CNM fraction as a function of metallicity*

In Figure 5 we show predictions from numerical simulations for how the CNM fraction depends on metallicity from the TIGRESS-NCR simulation suite (Kim et al. 2024). At low metallicity the CNM fraction is expected to decrease. While the dependence of cooling and heating on metallicity largely cancels out (within the metallicity range explored here), the CNM fraction depends subtly on the metallicity via UV radiation transfer and grain charging. Recent numerical simulations of the star-forming ISM, including an explicit UV radiation transfer and photochemistry (Kim et al. 2023), show that the reduced dust attenuation of FUV radiation at low metallicity makes photoelectric heating more efficient (Kim et al. 2024). The net effect is more reduction in cooling than heating, and as a consequence, a decrease in the CNM fraction. As shown in Figure 5, for their Solar neighborhood models, the resulting $f_{\rm CNM}$ drops to $\sim 0.1$ at $Z/Z_\odot = 0.3$ and $\sim 0.05$ at $Z/Z_\odot = 0.1$ (C.-G. Kim et al. in prep).

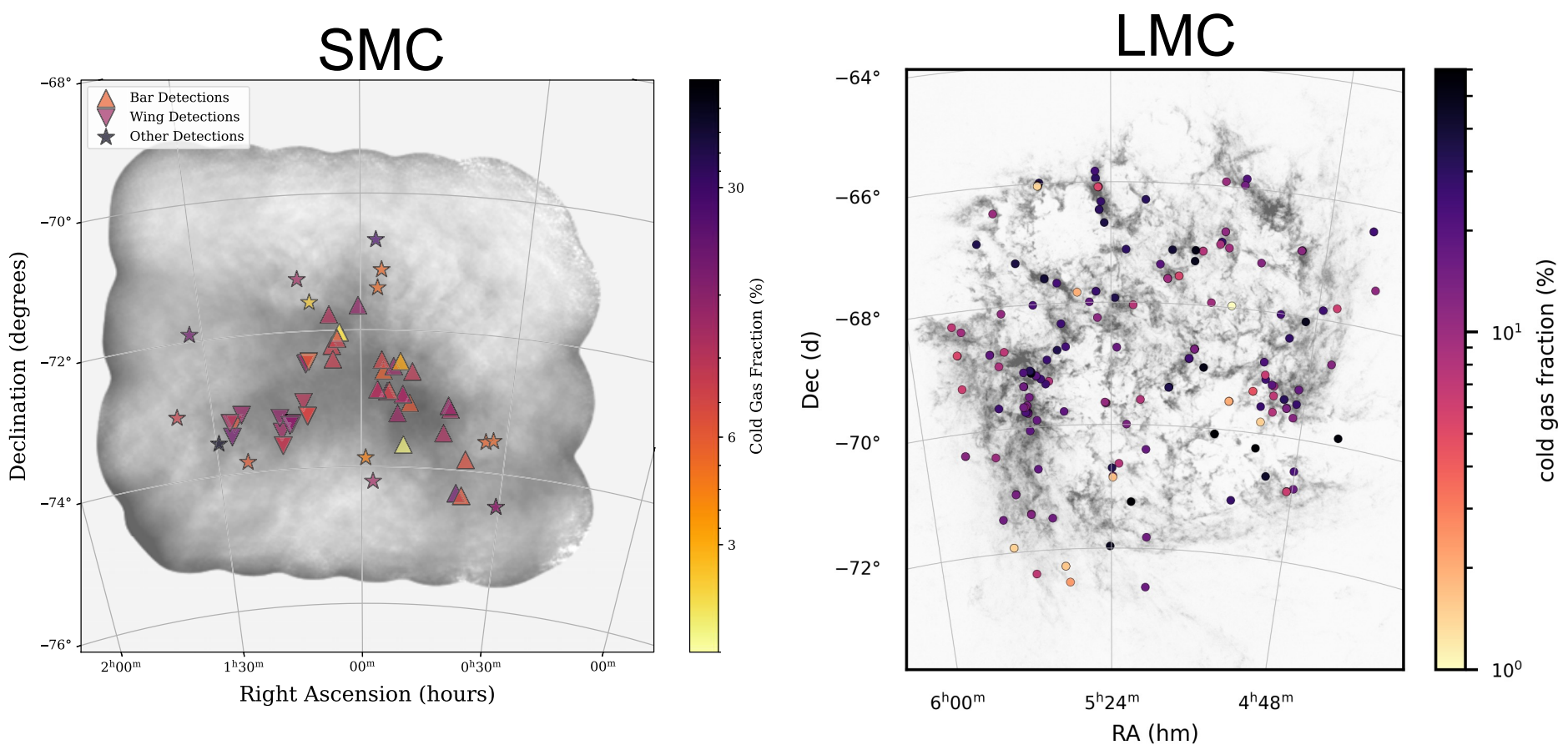

**Figure 4.** The CNM fraction distribution across the SMC from Dempsey et al. (2022) (left) and the LMC from Chen, Steffes, et al. in preparation (right). The CNM fraction was calculated under the assumption that the cold gas temperature is 30 K. Only sources where HI absorption was detected are shown.

Until recently, the only observational data point for comparison with these predictions was for the Milky Way. As we discussed in Section 4.1, there is a range of $f_{\rm CNM}$ values for the Milky Way that we show as error bars representing the 25th and 75th percentiles of the observed distributions. Recent large samples of HI absorption spectra for the Magellanic Clouds from GASKAP, shown in Figure 4, allow us to populate the sub-Solar regions in this figure. We show the median CNM fractions for the SMC from Dempsey et al. (2022) based on GASKAP observations. A preliminary analysis of GASKAP HI absorption spectra for the LMC from Chen, Steffes et al. (in preparation) is also shown. With future deeper HI observations of the Clouds by the GASKAP collaboration these constraints will become even firmer, although data so far suggest a good agreement between simulations and observations. While the point for the Milky Way agrees with simulations within provided ranges, the observations are reaching higher CNM fractions than what is seen in the simulations.

Finally, the ongoing Local Group L-Band Survey (LGLBS) is surveying six star-forming galaxies (M31, M33, WLM, NGC 6822, IC 10, IC 1613) in the Local Group using all configurations of the Very Large Array (Koch et al., in preparation). This survey is finding HI absorption in the direction of background as well as some internal radio continuum sources in these galaxies. As an example we show a data point for NGC 6822 from Pingel et al. (2024) based on a small sample of two sightlines.

These results demonstrate the power of high-resolution observations of nearby galaxies in probing the ISM theory under conditions different from what is found in the Milky Way. A lot more will come in the future from GASKAP and LGLBS.

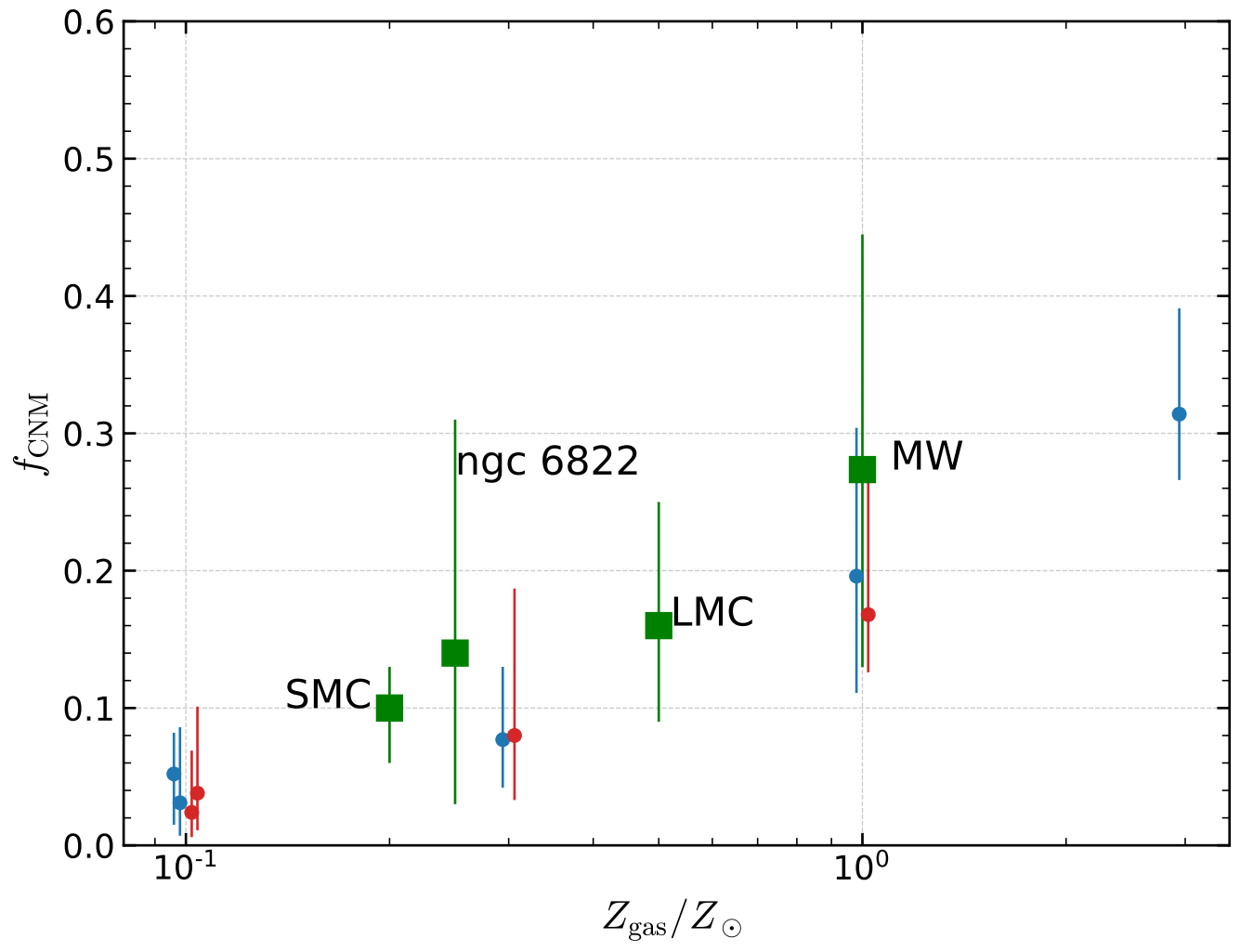

**Figure 5.** Red, blue circles: the predicted CNM fraction from Kim et al. (2024) TIGRESS-NCR simulations (Kim et al., private communication). Two different colors show two initial magnetic field values. Green squares show observational values from Dempsey et al. (2022), Pingel et al. (2024), and Chen et al, in preparation. The observed data points are median values calculated using all detected HI absorption spectra.

## 5. Thermally stable vs. unstable HI in the Milky Way

Measuring the mass fraction of HI as a function of temperature represents an important way of comparing observations with theory and numerical simulations. The simulations predict two clear peaks in the HI mass distribution, corresponding to the CNM and WNM, and some fraction of the thermally unstable HI (UNM). Over the last ten years numerical simulations have settled down to predicting 20-30% mass fraction of the UNM (some early studies found a substantial fraction of gas mass far from the thermal equilibrium values, e.g. Mac Low et al. (2005)). For example, Hill et al. (2018) found less than 20% of HI mass in the UNM phase when considering different heating functions, Kim et al. (2013) found $\sim 18\%$ in their 3D hydrodynamic simulations of galaxy disks with stellar feedback. Recent TIGRESS simulations by Kim et al. (2024) find $\sim 30\%$ in each of CNM and UNM phases, and $\sim 40\%$ in the WNM phase.

From the observational perspective, the BIGHICAT data set provides $T_s$ and the column density for individual velocity components detected in HI absorption spectra that can be used to calculate the HI mass fraction for different phases. This is shown in Figure 3 (left) with the black line. This figure shows the CNM peak of the HI mass distribution around 50-100 K, with a broad shoulder up to 500 K, and an extended tail up to $T_s = 3000$ K. However, the (absorption) observed HI gas fraction is missing a clear peak in the WNM portion of the distribution. Even with long integrations it has been very hard to detect the WNM in absorption.

We want to stress here that in interpreting the HI column density or mass distributions it is essential to understand survey completness and potential biases involved during the radiative transfer calculations. Synthetic spectra produced by numerical simulations provide an important way for calibrating observational methods. For example, Murray et al. (2017) used 10,000 synthetic HI emission-absorption pairs from the Kim et al. (2014) simulations to investigate how close their radiative transfer pipeline was in relation to the true excitation temperature. They were able to estimate the correction, or the transfer function, and apply it on the HI mass distribution. Similarly, by employing the Lindner et al. (2015) Gaussian decomposition technique, they showed that the recovery of individual HI structures is 99% for Galactic latitudes $|b| > 50^\circ$, while it drops to 53% for $|b| < 20^\circ$. Another example of using simulated data to test their method for $T_s$ estimation from HI absorption spectra was developed in Bhattacharjee et al. (2024).

### 5.1. *The fraction of the thermally unstable HI?*

The observational properties of HI at temperatures $T_s > 250$ K have been difficult to constrain. Until recently most estimates of the UNM temperature were made as upper limits from the line-width-based kinetic temperatures of either emission or absorption. For a summary of results please see McClure-Griffiths et al. (2023). Only a handful of observations have achieved the optical depth sensitivity required to measure spin temperatures of the UNM, e.g., Heiles and Troland (2003); Roy et al. (2013); Murray et al. (2018). The most robust estimates for the UNM fraction have come from the 21-SPONGE survey (Murray et al. 2018), which conducted very deep HI absorption observations with the key goal of measuring UNM and WNM properties. Not only that this was a large, statistically significant, sensitive survey (designed to detect low optical depth, broad lines), but this survey also took exceptional care of observational biases introduced in the analysis method used to constrain HI spin temperature and recover the HI mass distribution. The 21-SPONGE estimated that $\sim 20\%$ of the HI by mass is in the unstable phase. This result is similar to the HI emission derived UNM fractions (28% and 41%) by Marchal and Miville-Deschênes (2021) and Kalberla and Haud (2018) given the errors on those estimates, and in agreement with predictions from numerical simulations, e.g. TIGRESS (Kim et al. 2023, 2024).

However, there is still some debate regarding observational constraints on the UNM fraction in the literature. For example, studies by Roy et al. (2013) and Koley and Roy (2019), based on sensitive HI absorption spectra obtained with the WSRT and GRMT, found much higher fractions than most other observational studies. If taking into consideration all HI detected in absorption only, Koley & Roy (2019) found the following fractions of the CNM, UNM, and WNM: 15%, 75%, and 10%, respectively. This is drastically different from 21-SPONGE results: 56%, 41%, and 3%, respectively. Future work is needed to understand differences between different methodologies employed in these studies. Using the same definitions for temperature ranges for individual phases is also important for direct comparisons. With larger data sets of HI absorption spectra it will also be possible to distinguish whether some of the difference potentially stems from regional variations of UNM properties.

Finally, while until now measuring the HI mass distribution across different phases was possible only for the Milky Way, with large samples of HI absorption spectra for the SMC and LMC (especially with the upcoming deep ASKAP observations) we are in an excellent position to, for the first time, estimate the fractions of the CNM/UNM/WNM phases at low metallicity in nearby galaxies.

### 5.2. *The excitation temperature of the WNM?*

As the WNM usually dominates HI emission spectra, its properties are sometimes derived purely from HI emission data based on the width of fitted Gaussian functions, e.g. (Kalberla and Haud 2018; Marchal and Miville-Deschênes 2021). More robust estimates of the excitation temperature of the WNM, without a corruption by turbulent motions, come from using absorption-emission pairs. This approach requires highly sensitive interferometric HI absorption observations. For example, Carilli et al. (1998) measured $T_s = 5500 - 8700$ K in the direction of Cygnus A. Dwarakanath et al. (2002) and Kanekar et al. (2003) found lower values of 3000-3600 K. The 21-SPONGE project detected a handful of components with $T_s > 1000$ K (Murray et al. 2015). To improve sensitivity to shallow, broad absorption features further, Murray et al. (2014) employed a spectral stacking analysis on 1/3 of the survey data and detected a pervasive population of WNM gas with $T_s = 7200^{+1800}_{-1200}$ K. This result was confirmed using data from the entire 21-SPONGE survey in Murray et al. (2018), where residual absorption was stacked and binned based on residual emission, demonstrating the existence of a significant absorption feature with a harmonic mean $T_s \sim 10^4$ K.

The range of predicted kinetic temperatures for the WNM from the most detailed ISM heating and cooling considerations is $T_k \sim 5000 - 8800$ K (Wolfire et al. 2003). If the HI is only collisionally excited, this range implies $T_s \sim 1000 - 4000$ K (Kim et al. 2014). When, however, the common prescription is used to account for the Ly$\alpha$ flux and the Wouthuysen-Field effect, it is expected that $T_s \sim 3500 - 5000$ K. The Murray et al. (2014) and Carilli et al. (1998) results are significantly higher than predictions from standard ISM models based on collisional plus resonant Ly$\alpha$ excitation of the 21-cm transition.

As can be seen in Figure 3 (left), even all combined HI absorption measurements together are not enough to detect the WNM peak of the HI gas distribution. Therefore, Murray et al. (2018) suggested that the WNM spin temperature is likely higher than standard analytical and numerical predictions. Another issue could be the implementation of the $T_k$ to $T_s$ conversion in simulations and the use of a constant Ly$\alpha$ flux. In a recent study by Seon and Kim (2020), where a multi-phase model and simulations were used and photons originating from HII regions were tracked to produce the Ly$\alpha$ radiation field, the Ly$\alpha$ flux was strong enough to bring the 21-cm spin temperature of the WNM close to the kinetic temperature. This encouraging result demonstrates that a careful treatment of Ly$\alpha$ photons needs to be incorporated in numerical simulations, instead of the commonly used uniform Ly$\alpha$ flux.

### 5.3. *Spatial distribution of the CNM*

Finally, as the HI absorption measurements are limited by the position of background sources, the spatial distribution of the cold HI is largely unconstrained. As numerical simulations find a strong spatial correlation between $H_2$ and the CNM, observationally mapping the spatial distribution of the CNM leads to the finding chart of the future star formation sites. Rare examples of extended background radio continuum sources (e.g. supernova remnants) exist that allowed sampling of small-scale structure of the absorbing, cold HI, e.g. Deshpande et al. (2000). Another approach is HI self-absorption (HISA) which has been used to map the cold HI distribution in the Rigel-Crutcher cloud (McClure-Griffiths et al. 2006), or several nearby molecular clouds (Syed et al. 2023). Syed et al. (2023) found that the CNM mass traced by HISA around major molecular clouds is $\sim 3 - 9\%$ of the total HI mass, and that the cold HI has a widespread, yet clumpy and turbulent distribution. In the case of the Rigel-Crutcher cloud, the cold HI is organized in narrow filamentary structures with a typical width of $< 5$ pc and length $\sim 20$ pc. Large-scale filamentary distribution of the CNM was also revealed using statistical methods: Kalberla et al. (2016) found long (10-20$'$) filaments via Gaussian fitting of HI emission spectra, while Clark et al. (2014) applied the Rolling Hough Transform on the entire GALFA-HI data set and discovered many small (few arcminutes long) fibers whose narrow linewidth was suggestive of temperatures $< 200$ K. Often, cold HI filaments are found to be aligned with the ambient magnetic field orientation (Soler et al. 2022) and the connection between cold HI and the magnetic field is starting to be explored in the Magellanic Clouds (Ma et al. 2023).

There are two factors that will help with constraining the spatial distribution of the cold HI in the future. First, with wide-field surveys the number of background radio sources suitable for HI absorption measurements per square degree is drastically increasing. Denser grids of HI absorption will enable spatial interpolation and reconstruction of the cold HI distribution. Second, sophisticated numerical approaches are starting to be developed that can predict the CNM contribution from HI emission data. Murray et al. (2020) used HI absorption data to train a convolutional neural network (CNN) to predict the CNM fraction from HI emission data. Marchal et al. (2024) calculated Fourier transforms of HI emission spectra to extract narrow velocity components from all-sky HI surveys and estimate a lower limit on the cold

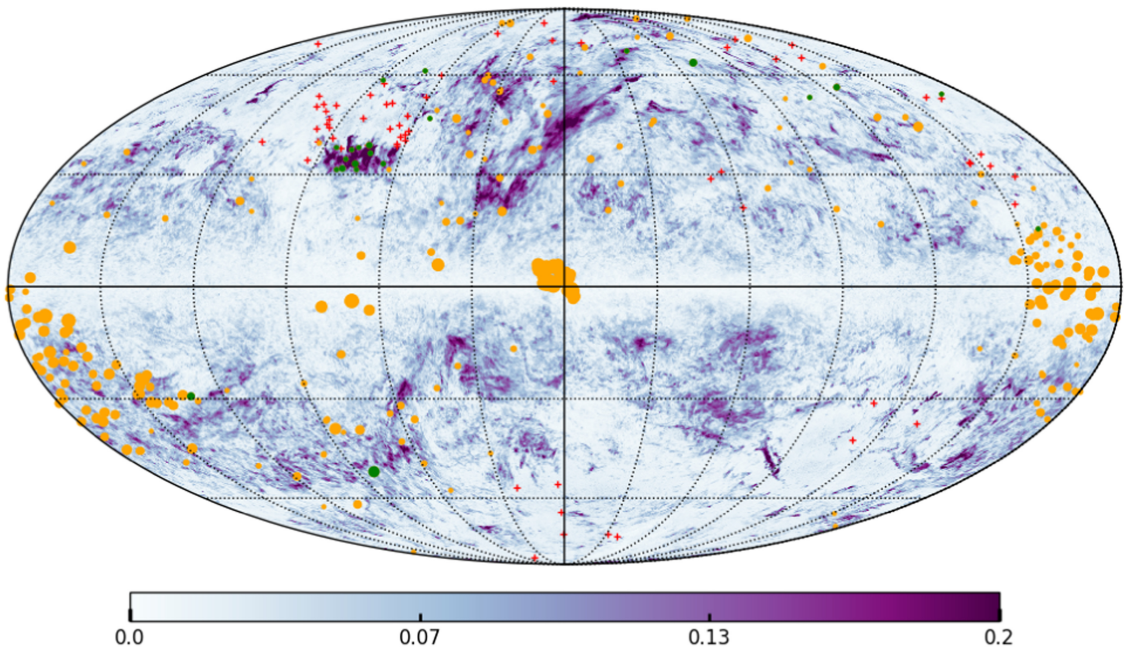

**Figure 6.** The lower limit on the cold gas mass fraction calculated for the entire HI4PI data set for the velocity range −90 to 90 km/s from Marchal et al. (2024). Data points shown are from the BIGHICAT and were used to compare the CNM fraction with the Fourier Transform results: red crosses are positions of non-detections, the orange points are detections with the symbol size encoding the number of absorption components (ranging from 1 to 16) in the optical depth spectra, the green dots show positions where the BIGHICAT estimate of the CNM fraction is smaller than that of the Fourier method.

HI mass fraction. We show their all-sky distribution of the CNM fraction in Figure 6. While all these studies have certain limitations and generally under-predict the CNM fraction, these approaches are highly encouraging. A combination of several different priors (e.g. from HI absorption, dust and/or linewidth) may be able to provide tighter constraints on the CNM properties and distribution in the future. Such approaches are especially important for external galaxies, where due to smaller solid angles, the HI absorption spectroscopy will be even more limited.

## 6. Conclusions

The SKA pathfinders (ASKAP, MeerKAT, VLA, and ngVLA in the future) are blurring the boundary between HI studies of the Milky Way and those of nearby galaxies. While understanding of detailed physical processes essential for setting the HI thermal balance was possible only in the Milky Way until recently, highly sensitive wide-field interferometric surveys are expanding our view into nearby galaxies with a diverse range of physical conditions. Future studies will provide essential tests of the ISM theory and chart the distribution and properties of the raw star formation fuel under metallicity and radiation field conditions that are vastly different from what we find in the Solar neighborhood. The SKA, once operational, will take the HI studies into a new era with dense grids of HI absorption for the Milky Way and nearby galaxies, and the ability to study detailed HI properties in galaxies beyond the Local Group. For all future observational studies, using synthetic data to calibrate analysis techniques and observational biases provides an essential way for estimating survey completeness and uncertainties of the measured physical parameters. We provide here a few takeaway points and future directions.

(1) The cold HI disk of the Milky Way is extended and coupled to molecular gas. The CNM is also colder and more abundant at higher optical extinction due to more intense cooling and shielding. With denser spatial sampling of HI absorption enabled by wide-field interferometric surveys, future studies will map spatial variations of the CNM properties within the Milky Way and nearby galaxies. Testing the $T_s$ and $f_{\rm CNM}$ dependence on $A_V$ in nearby galaxies will also be important.

(2) The extended parameter space offered by nearby galaxies is providing, for the first time, key tests of the HI excitation and thermodynamics. A comparison of the CNM fraction between the Milky Way, the SMC, the LMC and NGC 6822 shows a good agreement with predictions from numerical simulations (Kim et al. 2024) on how the CNM fraction depends on metallicity. But this is just the tip of the iceberg – future studies are needed to populate figures such as Figure 5 and establish relationships between properties of HI phases and local conditions, e.g. metallicity.
(3) The fraction and the spatial distribution of the thermally unstable HI remain as open questions, especially as tracing observationally the UNM is difficult even in the Milky Way. Large samples of HI absorption spectra can use a stacking analysis to probe the existence and fraction of the UNM in nearby galaxies.
(4) The excitation temperature of the WNM is not well understood. Observations suggest higher temperatures than what is seen in numerical simulations. More observational constraints are essential to understand environmental dependencies, especially in relation to the Ly$\alpha$ radiation field.
(5) Finally, while not discussed in this article, quantifying how stellar feedback affects HI turbulence is an important future goal. Wide-field interferometric surveys are expanding the spatial dynamic range for HI observations of nearby galaxies, which are essential for applying statistical methods and searching for local modifications of the turbulent power spectrum by individual feedback sources.

## Acknowledgments

SS is grateful to Chang-Goo Kim for providing simulation data for Figure 5. She would like to thank Chang-Goo Kim, Daniel Rybarczyk, Hongxing Chen and Nickolas Pingel for essential feedback and insightful discussions, and conference organizers for inviting her to this highly stimulating conference. This research was supported by the National Science Foundation award 2205630. Support for this research was also provided by the University of Wisconsin - Madison Office of the Vice Chancellor for Research and Graduate Education with funding from the Wisconsin Alumni Research Foundation.